\documentclass[aps,showpacs,onecolumn,floats,prd,superscriptaddress,nofootinbib]{revtex4} 
\usepackage{graphicx,amsmath,amssymb,amstext}
\usepackage{amssymb,amsbsy,amsfonts,amsthm,color}

\usepackage{slashbox}
\usepackage{epsfig}
\usepackage{graphicx}
\usepackage{subfigure}

\graphicspath{{Figures/}}

\begin{document}

\title{ Removing Skill Bias from Gaming Statistics }

\author{I-Sheng Yang}
\email{isheng.yang@gmail.com}
\affiliation{Canadian Institute of Theoretical Astrophysics, 60 St George St, Toronto, ON M5S 3H8, Canada.}
\affiliation{Perimeter Institute of Theoretical Physics, 31 Caroline Street North, Waterloo, ON N2L 2Y5, Canada.}

\begin{abstract}
``The chance to win given a certain move'' is an easily obtainable quantity from data and often quoted in gaming statistics.
It is also the fundamental quantity that reinforcement learning AI bases on.
Unfortunately, this conditional probability can be misleading.
Unless all players are equally skilled, this number does not tell us the intrinsic value of such move.
That is because conditioning on one good move also inevitably selects a subset of better players.
They tend to make other good moves, which also contribute to the extra winning chance.
We present a simple toy model to quantify this ``skill bias'' effect, and then propose a general method to remove it.
Our method is modular, generalizable, and also only requires easily obtainable quantities from data.
In particular, it gets the same answer independent of whether the data comes from a group of good or bad players.
This may help us to eventually break free from the conventional wisdom of ``learning from the experts'' and avoid the Group Thinking pitfall. 
\end{abstract}

\maketitle

\tableofcontents

\newpage

\section{Introduction}

In applied statistics, one major challenge that comes up all the time is to infer causation from correlation.
That is because data from any observation are inevitably biased by some selection effects.
A claim of causation based on observed correlation, no matter how natural it feels, may be an illusion.

One concrete example occurs in gaming statistics.
Human are obsessed with games.
We like to play, to watch, and most importantly, to win.
Thus, it is a common practice to collect records of past games and analyze which moves have been leading to victories.
One of the most natural statistics to take is {\bf the probably to win given a certain move}.
It is a well defined conditional probability, but quoting it can often be misleading.
When this quantity is quoted to a common folk, it sounds like the {\bf intrinsic value} of a move, which should have the following definition.
\begin{itemize}
\item Comparing ``making this move'' to ``not making this move'', how much more likely does a player win, {\bf given that all other conditions are the same.}
\end{itemize}

Unfortunately, this impression is seldom accurate, due to the fact that ``given that all other conditions are the same'' is almost never satisfied by the statistics.
In realistic data, not all participating players are equally skilled.
If we condition on a good move, then good players are more likely to make such a move.
They will also make other good moves, which all together increase the resulting winning chances.
Therefore, the conditional winning probability can be an over-estimation of the intrinsic value of that move.
Even more annoyingly, sometimes a bad move is favored by good players, and it can actually have a higher conditional winning probability just because of that.
This creates an misconception that such a bad move is actually good.

Sociological effects also compound this issue, which is sometimes referred to as ``group thinking'' in the gaming society.\cite{GroupThink}
When a game is complicated enough, people usually do not just learn the strategy from scratch.
We often learn from experts (good players), who will often tell you that the moves they make are good.
Most of those advices are actually correct, and your skill will improve as you start to follow them.
However, if those good players have any misconceptions, you will likely inherit them.
As you become a better player and your data shows up in statistics, the misconceptions are further solidified.

Of course, fundamentally, this is not a sociological issue.
It is a mathematical fact that
\begin{center}
(Conditional Probability to Win) $\neq$ (Intrinsic Value of a Move)
\end{center}
Treating them as being equal is nothing but a mathematical mistake.
It affects human players as well as sophisticate AIs.
AlphaGo Master and AlphaGo Zero have identical architecture, but the later is clearly superior than the former.
\cite{AlphaGoZero}
One major difference is that the former started learning from human experts, but the later did not.
This suggests that the initial human inputs did teach the AI some misconceptions, which the subsequent training from self-plays did not get rid off.
This is not entirely surprising, since reinforcement learning ultimately bases on the outcome of the game, which is effectively the conditional winning probability.
During the training process, the strategy of the AI evolves semi-stochastically, which means that the data is effectively taken from a group of players with a nonzero skill variation.
Thus, the resulting AI can suffer the skill-bias problem just like any human player.

In this paper, we provide a simple model to deal with this problem.
We propose a way to calculate the intrinsic value of a move based on statistical data which is still quite straightforward, with only 2 more quantities involved.
\begin{eqnarray}
{\rm Intrinsic \ Value} = \frac{ P_{cond.} - P_{skill} }{1 - 8\epsilon^2}~.
\label{eq-intro}
\end{eqnarray}
Here $P_{cond.}$ is the na\"ive conditional probability of winning given such move.
$P_{skill}$ is the probability for a random player from ``group 1'' to defeat a random player from ``group 2''. 
Group 1 consists of players with the same average skill as those who did make this move, and group 2 consists of players with the same average skill as those who did not make this move.
Subtracting $P_{skill}$ from $P_{cond.}$ is clearly understandable as removing any advantage from simply being better players.
Acute readers might immediately realize that such removal results in an underestimation.
Because ``making this move'' is part of ``being better players''.
That is why we need to divide the result by a factor that is slightly less than one.
$\epsilon$ is how much more likely does a player with skill 1-$\sigma$ higher than average to make such move, compared to an average player.
This formula is derived under the assumption that no single move changes the result dramatically.
Namely, $P_{cond.}$ and $P_{skill}$ should be close to $50\%$ for 2-player games, and $\epsilon \ll1$.
If these conditions are not ssatisfied, we cannot guarantee the accuracy of Eq.~(\ref{eq-intro}).

The two new quantities here, $P_{skill}$ and $\epsilon$, are easily computable from data if the number of games is much larger than (number of player)$^2$. 
When that is not the case, one can still use modern skill evaluation systems like True Skill \cite{TrueSkill} to derive the skill of each player from the data.
As long as such skill evaluation is reliable, we can use it to calculate  $P_{skill}$ and $\epsilon$.
Of course, this requires us to keep track of the players.
If the gaming data is player-blind, then our method is not applicable.
If no one plays the game repeatedly, then there is no way to have a reliable evaluation of their skills, and our method is also useless.

We should emphasize that other than the need to keep track of the players, our method does not require any extra data.
Actually, it is highly modular.
If we want to know the intrinsic value of one move, we still only need the data of that move.
Although fundamentally, the skill bias effect we want to remove is due to other moves, we managed to circumvent the need of using them.
Also, our method is independent from the average skill level of players.
We can learn the same intrinsic value from a group of experts or from a group of mediocre players.
This is only proven in our toy model where each move is mechanically independent from each other.
We are quite aware that such assumption may not be correct for realistic games.
Interestingly, applying our method to the actual data for a board game, this property seems to check out. \\ \ \\

{\bf Outline.} The rest of this paper goes like the following. \\ \ \\
In Sec.\ref{sec-toymodel}, we use a simple toy model to demonstrate the skill bias effect and to derive Eq.~(\ref{eq-intro}).
\ \\ \ \\
In Sec.\ref{sec-exp1}, we demonstrate how to compute Eq.~(\ref{eq-intro}) from basic statistical data. \\ \ \\
In Sec.\ref{sec-ToyExample}, we apply our method to the computerized version of our toy model to demonstrate two properties: (1) our result is independent of player skills, (2) our method can reveal important strategic bias. \\ \ \\
In Sec.\ref{sec-TTA}, we use the online scraped data of the board game: Through the Ages, to demonstrate the result of Eq.~(\ref{eq-intro}).
It surprisingly confirms the property that the intrinsic values obtained by our method are independent of the average skill of players in the data.
It also reveals real misconceptions that exist among the online players of this game.

\section{Skill Bias Removal: Derivation from a Toy Model}
\label{sec-toymodel}

We will start by setting up an abstract, 2-player, 0-sum, not entirely deterministic game with no ties.
The game consists of $N$ binary matches.
Real numbers $c_n$ for $n=1\sim N$ quantify the intrinsic values of these matches.
$d_n = 1$ or $-1$ means that whether the outcome of the $n$th match is ``favorable'' to either player $A$ or $B$ (if $c_n>0$).
By definition, when a match outcome is favorable to one player, it must be unfavorable to the other player.
The probability for player $A$ to win is given by
\begin{equation}
P_A = 0.5 + \sum_n d_n c_n~.
\label{eq-game}
\end{equation}
Naturally, for player $B$, it is
\begin{equation}
P_B = 0.5 - \sum_n d_n c_n~.
\end{equation}
We will assume that $c_n$'s are small enough such that the probabilities are bounded between 0 and 1.\footnote{
Strictly speaking, we should use a sigmoid or $\arctan$ function to achieve that.
However, for the purpose of this paper, we will skip such nonlinear technicalities.
When the value of $\sum_n d_n c_n$ stays small within the approximately linear response range of those functions, our result is valid.
Situations very different from such regime will be a topic of future work.}
All these parameters are actually hidden.
They will help us to visualize the problem and find a solution, while the solution will be independent of these parameters.

In this game, a ``move'' is basically a successful attempt to secure the ``favorable'' outcome of a match.
A ``good move'' would be securing a match whose $c_n$ is indeed larger than zero.
Thus, we will model the strategy of a player by $N$ numbers, $w_n$.
For player $A$, we denote these numbers by $w_n^A$, which basically means how much player $A$ wants to make the outcome of the $n$th match ``favorable'' to him.
The probability for that to actually happen to player $A$ against player $B$ is
\begin{equation}
P(d_n=1) = 0.5 + w_n^A - w_n^B~.
\label{eq-match}
\end{equation}
One may ask why would someone ``not want'' the favorable outcome.
Well, the thing is that, the game is complicated enough.
No players are absolutely certain about the sign of each $c_n$.
For all they know, $c_n$ might be negative.
In that case, a small $w_n$ would have been the better strategy.

The ``skill'' of a player is quantified in the following way.
Let $s^A$ be the skill of player $A$, the strategy $w_n^A$ is given as the sum of two terms.
\begin{equation}
w_n^A = \lambda_n s^A + b_n^A~.
\end{equation}
$b_n^A$'s are independent random numbers with zero mean.
\begin{eqnarray}
\langle b_n^A \rangle &=& 0~, \\
\langle b_n^A b_m^B \rangle &\propto& \delta_{AB} \delta_{mn}~.
\end{eqnarray}
Their existence guarantees that no 2 players are identical.
As we will see, they will play no role in our result.
We have the freedom to choose any variation of $b_n$ per player per match and still get the same result.
That speaks for the generality of our method.

$s^A$ is the ``skill'' of player $A$, which is the tendency to consistently get more favorable outcomes.
If $\lambda_n c_n>0$, then $|\lambda_n|$ is the ``simplicity'' of match $n$.
Larger $|\lambda_n|$ implies a simple choice---only a small skill advantage is needed to recognize which outcome is actually favorable in the $n$th match.
On the other hand, $\lambda_n c_n<0$ implies a ``misleading'' choice---better players are actually less likely get the favorable outcome in this match.

The expected chance for someone with skill $s^A$ to defeat someone with skill $s^B$ is given by
\begin{eqnarray}
\langle P(A {\rm \ defeats \ } B) \rangle &=&
0.5 + \sum_n \langle d_n\rangle c_n~, \\
\langle d_n \rangle &=& P(d_n=1) - P(d_n=-1) = 
2 ( w_n^A - w_n^B )~,
\\
\langle P(A {\rm \ defeats \ } B) \rangle &=& 0.5 + 2 \sum_n \langle w_n^A - w_n^B \rangle c_n
=  0.5 + 2 (s^A - s^B) \sum_n \lambda_n c_n~.
\end{eqnarray}
Thus, without loss of generality, we demand that $\sum_n \lambda_n c_n > 0$.
This ensures that more skilled players indeed have higher winning chance.
\footnote{If this were not the case, simply flip the sign of all $\lambda_n$.}

Now, assume that there are $M\gg1$ players, and their skill distribution is
\begin{equation}
\langle s \rangle =S~, \ \ \ \langle (s^A-S) (s^B-S) \rangle = \delta_{AB} \sigma_s^2~.
\end{equation}
From the data of many games between these players, we would like to ask the following question: \\ \ \\
{\bf How do we find out the importance of a particular match $c_n$?} \\ \ \\
Na\"ively, we should look for the conditional probability that ``how likely for a player to win if the $n$th outcome is favorable.'' 
This however, is only correct when all players have the same skill, $\sigma_s \sim 0$.
Because while conditioning on the outcome of the $n$th match, we are also selecting a slightly more skilled subset of players.
The increased winning probability also comes from the fact that their superior skills tend to swing other matches in their favor.
This is what we call the ``skill bias'' effect.

Here is the math.
\begin{eqnarray}
P\left( A {\rm \ wins} \bigg| d_n=1 \right) &=& 
\label{eq-condition}
\frac{ \left\langle \left(0.5 + w_n^A - w_n^B  \right) \left( 0.5 + c_n +  \sum_{m\neq n} d_m c_m \right) \right\rangle }
{ \langle 0.5 + w_n^A - w_n^B \rangle }
\\ \nonumber
&=& 2*\bigg[ 0.25 + 0.5c_n + 
\sum_{m\neq n} 2 c_m 
\left\langle \left(w_n^A - w_n^B\right) \left( w_m^A - w_m^B \right) \right\rangle \bigg]
\\ \nonumber
&=& 0.5 + c_n + 8\lambda_n \sigma_s^2 \sum_{m\neq n} \lambda_m c_m~.
\end{eqnarray}
This clearly shows the two contributions:
the real importance of the match $c_n$, and the extra contribution through other matches which is the skill bias.
Thus, in order to find $c_n$, we need to know the value of this skill bias term.

We can first ask about the average skill for players who get (un)favorable outcomes from match $n$.
\begin{equation}
\langle s -S\rangle_{d_n=\pm1} = 
\frac{\left\langle \left(0.5 \pm w_n^A \mp w_n^B  \right) (s^A-S)\right\rangle}
{\left\langle \left(0.5 \pm w_n^A \mp w_n^B  \right) \right\rangle}
= \pm 2\lambda_n \sigma_s^2~.
\label{eq-skill}
\end{equation}
As a reality check, let us think about the situation when $\lambda_n\rightarrow0$.
This means that match $n$ is quite nontrivial, such that even good players cannot determine which outcome is favorable.
Naturally, everyone chooses randomly, and conditioning on such choice does not lead to a skill bias.

Next, for the two subsets of players with average skill given by Eq.~(\ref{eq-skill}), we calculate the expected chance that a random player form one subset defeats someone from the other subset.
\begin{equation}
P\left( A {\rm \ wins} \bigg| \frac{ \langle s^A\rangle =  \langle s \rangle_{d_n=1} }{ \langle s^B\rangle =  \langle s \rangle_{d_n=-1} } \right) 
= 0.5+ 8\lambda_n \sigma_s^2 \sum_m \lambda_m c_m~.
\label{eq-subset}
\end{equation}
Combine Eq.~(\ref{eq-condition}) and (\ref{eq-subset}), we get the following formula very close to the final result. 
\begin{eqnarray}
(1 - 8\lambda_n^2\sigma_s^2)c_n = P\left( A {\rm \ wins} \bigg| d_n=1 \right) -
 P\left( A {\rm \ wins} \bigg| \frac{ \langle s^A\rangle =  \langle s \rangle_{d_n=1} }{ \langle s^B\rangle =  \langle s \rangle_{d_n=-1} } \right)
\label{eq-cn}
\end{eqnarray}

The final input we need is the standard deviation in skill.
$\sigma_s$ is just a theoretical, unobservable parameter in our model.
The corresponding observable is the standard deviation in winning chances.
\begin{eqnarray}
\left\langle \left[ P(A {\rm \ defeats \ } B) - 0.5 \right]^2 \right\rangle
= 8 \sigma_s^2 \left( \sum_n \lambda_n c_n \right)^2~.
\label{eq-spread}
\end{eqnarray}
Combining Eq.~(\ref{eq-spread}) and (\ref{eq-subset}), we get
\begin{equation}
\lambda_n \sigma_s = 
\frac{P\left( A {\rm \ wins} \bigg| \frac{ \langle s^A\rangle =  \langle s \rangle_{d_n=1} }{ \langle s^B\rangle =  \langle s \rangle_{d_n=-1} } \right)   - 0.5 }
{\sqrt{ 8 \left\langle \left[ P(A {\rm \ defeats \ } B) - 0.5 \right]^2 \right\rangle} }
\label{eq-lambda}
\end{equation}

Combine Eq.~(\ref{eq-lambda}) and Eq.~(\ref{eq-cn}), we can express $c_n$ entirely with observable quantities from data.
\begin{eqnarray}
c_n = \frac{P\left( A {\rm \ wins} \bigg| d_n=1 \right) -
 P\left( A {\rm \ wins} \bigg| \frac{ \langle s^A\rangle =  \langle s \rangle_{d_n=1} }{ \langle s^B\rangle =  \langle s \rangle_{d_n=-1} } \right)}
{1 - 8 \left[ \frac{P\left( A {\rm \ wins} \bigg| \frac{ \langle s^A\rangle =  \langle s \rangle_{d_n=1} }{ \langle s^B\rangle =  \langle s \rangle_{d_n=-1} } \right)   - 0.5 }
{\sqrt{ 8 \left\langle \left[ P(A {\rm \ defeats \ } B) - 0.5 \right]^2 \right\rangle} } \right]^2}~.
\label{eq-result}
\end{eqnarray}
This is basically Eq.~(\ref{eq-intro}) in a more technical form.

\section{How does it work?}
\label{sec-exp1}

Eq.~(\ref{eq-result}) contains three probabilities that we should read from data.
Among them, $P\left( A {\rm \ wins} \bigg| \frac{ \langle s^A\rangle =  \langle s \rangle_{d_n=1} }{ \langle s^B\rangle =  \langle s \rangle_{d_n=-1} } \right)$ is somewhat a new idea.
Let us look at the following example.
\begin{center}
\begin{tabular}{ | c | c | c | c | c | c |  }
\hline
      & Game 1 & Game 2 & Game 3 & Game 4 & Game 5 \\ 
\hline
$d_n = 1$ & A & C & C & c & b \\  
\hline
$d_n =-1$ & b & b & a & B & A  \\
\hline
\end{tabular} $\times 20$
\end{center}
This is the record of 100 games between 3 players, A, B, and C.
Capital letter means that the player wins the game.
The first row is the player who gets the favorable outcome from the $n$th match.
For simplicity, we assume that every 5 games are exactly the same, thus the above table is repeated 20 times to form the entire data set.

Clearly, we have
\begin{equation}
P\left( A {\rm \ wins} \bigg| d_n=1 \right) = 0.6~,
\end{equation}
which seems to suggest that $d_n=1$ is the favorable outcome in this match.
However, we can see that $A$ and $C$ seems to be better players both with $0.67$ winning chances, and $B$ is a weaker player with only $0.25$ winning chance.
Incidentally, players $A$ and $C$ contribute to all the wins for $d_n=1$, while player $B$ contributes to the majority of losses for $d_n=-1$.
It is not directly clear whether $d_n=1$ is truly favorable, or simply because it is the prefered choice of the better players.

In order to resolve this ambiguity, our method requires us to calculate $P\left( A {\rm \ wins} \bigg| \frac{ \langle s^A\rangle =  \langle s \rangle_{d_n=1} }{ \langle s^B\rangle =  \langle s \rangle_{d_n=-1} } \right)$.

One intuitive approach is to derive it directly from pair-wise winning chances in the data.
\begin{eqnarray}
P_{AB} = 1-P_{BA} = P(A {\rm \ defeats \ } B) &=& 1~, \nonumber \\
P_{BC} = 1-P_{CB} = P(B {\rm \ defeats \ } C) &=& 0.5~, 
\label{eq-direct}
\\ \nonumber
P_{CA} = 1-P_{AC} = P(C {\rm \ defeats \ } A) &=& 1~.   
\end{eqnarray}
Also, by definition, $P_{AA}=P_{BB}=P_{CC}=0.5$.
\footnote{Note that two of the pair-wise winning chances are exactly $100\%$, which is somewhat outside the scope of our assumptions of the previous section.
So we should not talk the result too seriously.
This is just a toy example to demonstrate how our method works.}

From these probabilities, we can make Table (\ref{tb-1}).
\begin{table}[h!]
\begin{tabular}{ | c | c | c | c | c | c |  }
\hline
\backslashbox{$d_n=-1$}{$d_n=1$}      & A & C & C & C & B \\ 
\hline
B & $P_{AB}$ & $P_{CB}$ & $P_{CB}$  & $P_{CB}$  & $P_{BB}$ \\  
\hline
B & $P_{AB}$ & $P_{CB}$ & $P_{CB}$  & $P_{CB}$  & $P_{BB}$ \\  
\hline
A & $P_{AA}$ & $P_{CA}$ & $P_{CA}$  & $P_{CA}$  & $P_{BA}$ \\  
\hline
B & $P_{AB}$ & $P_{CB}$ & $P_{CB}$  & $P_{CB}$  & $P_{BB}$ \\  
\hline
A & $P_{AA}$ & $P_{CA}$ & $P_{CA}$  & $P_{CA}$  & $P_{BA}$ \\  
\hline
\end{tabular}
\caption{The table for calculating $P\left( A {\rm \ wins} \bigg| \frac{ \langle s^A\rangle =  \langle s \rangle_{d_n=1} }{ \langle s^B\rangle =  \langle s \rangle_{d_n=-1} } \right)$.}
\label{tb-1}
\end{table}
Then we can plug in Eq.~(\ref{eq-direct}), add up all those numbers in the table, and divide by 25.
That gives us
\begin{equation}
P\left( A {\rm \ wins} \bigg| \frac{ \langle s^A\rangle =  \langle s \rangle_{d_n=1} }{ \langle s^B\rangle =  \langle s \rangle_{d_n=-1} } \right) = 0.64~.
\end{equation}
Since this is actually higher than $P\left( A {\rm \ wins} \bigg| d_n=1 \right) = 0.6$, it suggests that $c_n<0$ and $d_n=1$ is not really a favorable outcome of match $n$.

In reality when the number of games is large, we may not really want to calculate the entire table.
We can sample the table sparsely by randomly selecting players from the $d_n=\pm1$ groups and match them with each other.
One thing to note here is that in our method, we are always weighting by games, not players.
Thus the random sample from the $d_n=1$ subset should be drawn from \{A:1, B:1, C:3\} and allow repeats.
Namely, a player who appears in more games will contribute more to the probability.
That is because the calculation of Eq.~(\ref{eq-condition}) naturally requires such weighting, and all other expectation values must follow the same rule to be self-consistent.
In particular, this weighting guarantees that the average winning chance is $50\%$. 
That would not have been the case if we weight by players.
Finally, $\langle (P-0.5)^2 \rangle$ can be calculated by randomly drawing pairs of players from \{A:3, B:4, C:3\}, which is always weighted by their appearance in all recorded games.

The reason why we want this example to have 100 games instead of 5 is that we want to take the probabilities in Eq.~(\ref{eq-direct}) seriously.
If there were only 5 games, we would have been claiming three things:
\begin{itemize}
\item $A$ will defeat $B$ all the time, based on the fact that it happened twice.
\item $C$ will defeat $A$ all the time, based on the fact that it happened once.
\item $B$ will defeat $C$ $50\%$ of the time, based on two results, and ignoring their relative performance against $A$.
\end{itemize}
We can see that these are not the best assumptions.
However, if $A$ indeed defeats $B$ 40 times in a row, the actual winning chance might be really close to $100\%$.
Thus, if (number of games)$\gg$(number of players)$^2$, then we would have enough statistics for the majority of pair-wise winning chances.
In those situations, using Eq.~(\ref{eq-direct}) is fine.

When the number of games is not that large, not only the statistical winning percentage is questionable, games between certain pairs of players might simply do not exist.
In those cases, we should use functions like True Skill \cite{TrueSkill}, which will give us better estimations on the pair-wise winning chances.
Of course, even the True Skill system requires enough number of games per player, and each player should at least be indirectly connected.
{\bf Thus, the limitation of applying our method is the requirement that the skill of majority of players can be reliably evaluated.}

\section{A Simple Example}
\label{sec-ToyExample}

We have written the game described in Sec.\ref{sec-toymodel} into a computer code, with the nonlinear modifications to Eq.~(\ref{eq-match}) and (\ref{eq-game}).
\begin{eqnarray}
P_A &=& 0.5 + \frac{\arctan(\pi \sum c_n d_n)}{\pi}~, \\
P(d_n=1) &=& 0.5 + \frac{\arctan(w^A_n - w^B_n)}{\pi}~.
\end{eqnarray}
When the arguments of $\arctan$ are small, nonlinearities are not important, and it should behave it the same way as we calculated analytically.
The players are coded similarly, with the extra constraint that their strategies, $w_n$, are $N$-dimensional vectors normalized to a given length, $(\Omega\sqrt{N})$.
The normalization is imposed to prevent the values of $w_n$ from running away during a training process, and for the convenience that the typical value of $|w_n|$ is about $\Omega$.
\footnote{Due to this normalization, the statistics of $w_n$ will not be exactly the same as in Sec.\ref{sec-toymodel}.
That is not a problem since one of our goals is to check whether our method is applicable generally, without the specific assumption about the player skills.}
This can also be understood as part of the game rules that all players have the same finite total budgets that they can use to compete during each match.

\begin{figure}[tb]
\includegraphics[width = 0.5\textwidth]{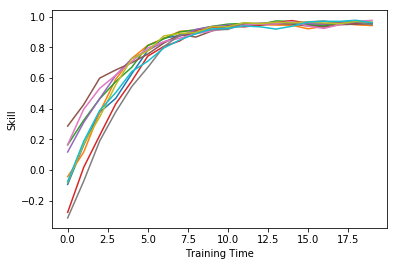}
\caption{The skill improvement of 10 players through reinforcement training.}
\label{fig-training}
\end{figure}

Clearly, if we treat both the lists of $c_n$ and $w_n$ as vectors, when the nonlinearities are not important, the best strategy should be pretty close to satisfying $\frac{\vec{c}\cdot\vec{w}}{|c||w|}=1$.
In fact, we will use this inner product as our way to evaluate skill, and a little reinforcement learning can confirm that it is a good choice.
We started 10 players with random vectors $w_n^{random}$, and allowed them to play against each other, modified their strategies according to the outcome of each game.
We plot their values of $\frac{\vec{c}\cdot\vec{w}}{|c||w|}$ in Figure \ref{fig-training} and see that they in indeed grow and approach 1.

For better visualization, we will focus on a game with $N=30$ and $c_n=1\%$ for all $n$.
Namely, securing every single match will increase your winning chance for about $1\%$ (can be less if you are far away from $50\%$ due to the nonlinearities).
Obviously, the best strategy is to have all $w_n$ positive and equal.
We then introduce a ``misguided expert''.
This player follows a strategy that is quite good.
\begin{equation}
w_0^{exp} = -\Omega~, \ \ \ w_n^{exp} = \Omega~, \ \ \ \forall~n\neq0~.
\end{equation}
Basically, he got all matches but 1 right.
Although, the one he got wrong, he got that completely wrong.

This misguided expert is not playing himself.
He is well respected and people learn from him.
We make another 10 players who start with random strategies, but instead of reinforcement learning, we modify their strategies in the following ways to mimic ``learning'' from this expert.
\begin{equation}
w_n = \lambda w_n^{exp} + w_n^{random}~, \ \ \ {\rm renormalized},
\label{eq-expert}
\end{equation}
with the value of $\lambda$ varies from one player to another.

\begin{figure}[tb]
\begin{minipage}{.5\textwidth}
\includegraphics[width = 0.9\linewidth]{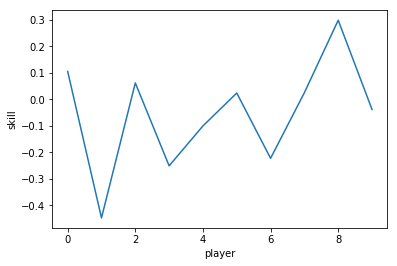}
\end{minipage}%
\begin{minipage}{.5\textwidth}
\includegraphics[width = 0.9\linewidth]{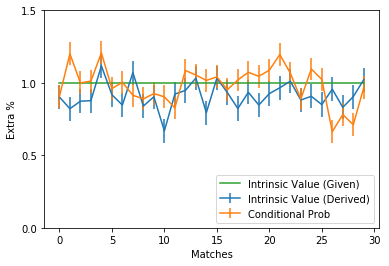}
\end{minipage}

\begin{minipage}{.5\textwidth}
\includegraphics[width = 0.9\linewidth]{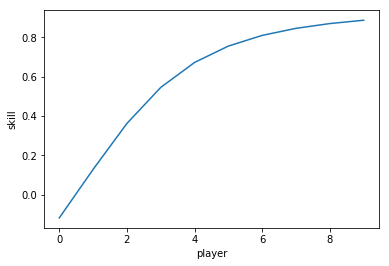}
\end{minipage}%
\begin{minipage}{.5\textwidth}
\includegraphics[width = 0.9\linewidth]{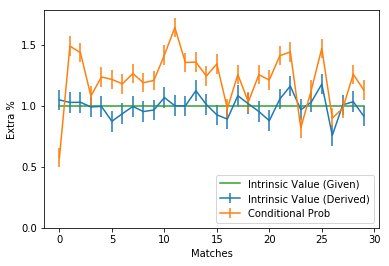}
\end{minipage}
\caption{{\bf Left:} The skill distribution among the 10 players.
{\bf Right:} Comparing the conditional winning probabilities, the actual values of $c_n$, and the values of $c_n$ recovered by our method.
{\bf Top:} Players with random strategies.
{\bf Bottom:} Players affected by the misguided expert up to various degrees.  Note that the misguided expert is wrong about $c_0$, and the conditional probability is significantly affected by that. The skill-bias removal brings $c_0$ back to the value it should be, $1\%$. }
\label{fig-Top}
\end{figure}

In the following 2 examples, we have $\Omega=10\%$.
Nonlinearities will be somewhat significant in each match, but we need to live with that to see the effect.
Namely, we need to make sure that players have strong opinions on each match, despite the fact that their individual influence on the final result is small.
This, fortunately, is quite a common behavior for human players.
In Figures \ref{fig-Top}, we compare the results between a group of 10 players with random strategies, and 10 players who learned form the misguided expert to various degrees.
We can see that the conditional winning probabilities are strongly influenced by the strongly biased teaching from the misguided expert.
$c_0$, the match that the misguided expert considers to be bad, does perform much poorly in terms of the conditional winning probability.
At the same time, other matches, for which the misguided expert's opinion is correct, perform better than they should.
Our method of skill-bias removal always recovers the value of $c_n=1\%$ more accurately.
Even among random players, our answer seems to track the actual values of $c_n$ slightly better than the conditional probability.

On top of recovering the value of $c_n$ more accurately, these 2 examples also supports our claim that the method works independent of the average skill level of the players.
The expert, though misguided, does teach mostly correct lessons.
Thus, the group of players to learned from him are indeed better, and we recover $c_n$ from the records of both groups equally well.

This example helps us to demonstrate the skill-invariance of our method, and its advantage over using the conditional winning probability directly.
However, one may wonder how much these 2 properties remain true in more realistic situations.
In particular, 
\begin{itemize}
\item In our toy model, every match is mechanically independent from one another (except for the common budget implies by the normalization of $w_n$.
In a realistic game, almost every decision are intricately connected.
Will the skill-invariance of our method remain true?
\item Skill-bias removal seems to be most important when there is a strongly biased opinion among the players, such as the one introduced by our imaginary, misguided expert.
How often is that true?
\end{itemize}
In the next section, we will demonstrate the result with an actual board game with only scraped data, which may shed some lights on these 2 concerns.

\section{A Realistic Example}
\label{sec-TTA}

\textbf{Through the Ages: A New Story of Civilization} is a deep strategic boardgame for 2-4 players.
There is a website that allows real people to play against each other, and it keeps records of past games.
We scraped the data of 30k+ games and use them as our example.

First of all, the multi-player nature has a well-known solution.
A game with more than 2 players will be treated as multiple 2-player games by comparing the results between all pairs present.
Next, any non-binary decision can be decomposed into multiple binary decisions.
For example, in this game, each player can try to play cards from a common pool, which are limited in supply.
Instead of treating all cards together like a complicate multi-choice match, we can treat each cards separately.
Each card is treated as a match with 2 outcomes---whether you play it or not.
Then, between two players who chose differently in the same game, we can apply our method.

For example, in a game with 4 players, $A$, $B$, $C$ and $D$, who ended up with scores $94$, $88$, $130$, $110$.
If player $D$ is the only one who played a certain card, then we treat it as 3 games.
\begin{center}
\begin{tabular}{ | c | c | c | c | c }
\hline
      & Game 1 & Game 2 & Game 3 \\ 
\hline
$d_n = 1$ & D & D & d \\  
\hline
$d_n =-1$ & a & b & C   \\
\hline
\end{tabular}
\end{center}
Note that player $C$ does defeat $A$ and $B$.
However, those are only taken into account in the computation of their True Skill.
Since none of them played this certain card, the result of this game between those players are irrelevant in its evaluation.
We applied this method to all games, then we obtained the value of the card $c_n$.

\begin{figure}[tb]
\includegraphics[width = 0.5\textwidth]{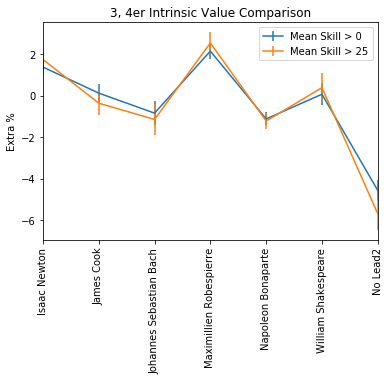}
\caption{The intrinsic value of each Age II Leader after removing skill bias effect.
Blue curve is obtained from using all game data.
Orange curve only uses data from games where average player skills are in the higher $50\%$.}
\label{fig-TTA}
\end{figure}

In Figure \ref{fig-TTA}, we show the true value of all Leader cards from Age II of the game, including the value of not choosing a leader at all.
First of all, the result is not far from common consensus among seasoned players.
When applied to all cards in the game, we do get a few unexpected results.
But at least 90\% of the results will not raise strong objections.

Another interesting thing to see in Figure \ref{fig-TTA} is that we tried to test how the result changes based on the average skill of players we collect data from.
According to our toy model in Sec.\ref{sec-toymodel}, the result should be invariant, and it indeed is.
That is a little surprising.
Every card (match) in our toy model is mechanically independent from each other.
It is natural to suspect that such simplifying assumption is essential to the apparent independence from average player skill.
In Through the Ages, in particular, in Figure \ref{fig-TTA}, the decision of taking each card here is clearly not independent from each other.
These cards will become available in different sequences from game to game.
They stay available and useful for different durations, and taking one of them forbids you from taking another.
Thus, in a multi-player game, the decision to take each card is highly intertwined.
Not to mention that each card may open up a different set of other choices, thus their next effect is likely highly nonlinear and profound.
However, when the effect of each decision is subtle enough that no one can be absolutely certain, treating them as being independent does not seem to be a bad assumption.

\begin{figure}[tb]
\includegraphics[width = 0.5\textwidth]{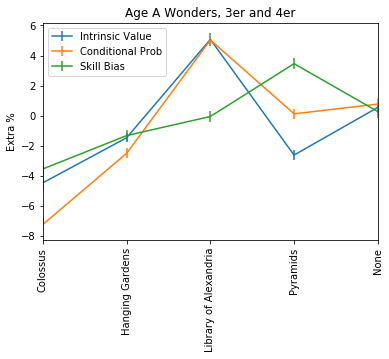}
\caption{The importance of each Age A Wonder after removing Skill Bias.
Blue curve is the intrinsic value after moving skill bias effect.
Orange curve is the conditional winning probability.
Green curve is the preference of good players, $\epsilon=\lambda_n\sigma_s$, namely, the extra probability that someone with skill 1-$\sigma$ higher than average will choose that card.
 }
\label{fig-TTA1}
\end{figure}

Let us go back to the few unexpected results.
Figure \ref{fig-TTA1} shows the intrinsic value of each Age A Wonder cards.
We choose this set of cards because it demonstrate a clear misconception, almost similar to the ``teaching from a biased expert'' as we designed in the previous section.
We can see that the card ``Pyramids'' is most favored by good players, but its intrinsic value is not the highest.
In fact, its conditional winning probability, $P(A {\rm \ wins}|d_n=1)$, is consistent with only $50\%$.
However, because good players prefer it so much, 
$P(A {\rm \ wins}|\frac{\langle s^A\rangle = \langle s\rangle_{d_n=1}}{\langle s^B\rangle = \langle s\rangle_{d_n=-1}}) > P(A {\rm \ wins}|d_n=1)$, the removal of skill bias effect revealed that it is actually a bad card.
In the technical terms of our toy model in Sec.\ref{sec-toymodel}, Pyramids have $c_n<0$ yet $\lambda_n>0$, thus it is a misleading choice---good players evaluate such card incorrectly.
In fact, we can see that another card here, Library of Alexandria, has $c_n>0$ yet $\lambda_n$ consistent with being 0.
That implies a difficult choice---although it is actually a good card, good players failed to recognize that.
At least, better players do not pick this card more often than worse players.

We are pleasantly surprised by both facts:
\begin{itemize}
\item The skill-invariance of our method persists in a realistic game and data.
\item Our method does differ significantly from using conditional probability only, and it demonstrate important misconceptions in the strategy of this game.
\end{itemize}
We hope to see our method applied to more games and data.

\acknowledgments

We thank the administrator of boardgaming-online.com for tolerating the scraping of data used in this paper. 
We also thank the players from boardgamegeek.com for interesting discussions.

\bibliography{all_active}

\begin{thebibliography}{3}
\expandafter\ifx\csname natexlab\endcsname\relax\def\natexlab#1{#1}\fi
\expandafter\ifx\csname bibnamefont\endcsname\relax
  \def\bibnamefont#1{#1}\fi
\expandafter\ifx\csname bibfnamefont\endcsname\relax
  \def\bibfnamefont#1{#1}\fi
\expandafter\ifx\csname citenamefont\endcsname\relax
  \def\citenamefont#1{#1}\fi
\expandafter\ifx\csname url\endcsname\relax
  \def\url#1{\texttt{#1}}\fi
\expandafter\ifx\csname urlprefix\endcsname\relax\def\urlprefix{URL }\fi
\providecommand{\bibinfo}[2]{#2}
\providecommand{\eprint}[2][]{\url{#2}}

\bibitem[{\citenamefont{Turner and Pratkanis}(1998)}]{GroupThink}
\bibinfo{author}{\bibfnamefont{M.~E.} \bibnamefont{Turner}} \bibnamefont{and}
  \bibinfo{author}{\bibfnamefont{A.~R.} \bibnamefont{Pratkanis}},
  \bibinfo{journal}{Organizational Behavior and Human Decision Processes}
  \textbf{\bibinfo{volume}{73}}, \bibinfo{pages}{105 } (\bibinfo{year}{1998}),
  ISSN \bibinfo{issn}{0749-5978},
  \urlprefix\url{http://www.sciencedirect.com/science/article/pii/S074959789892756X}.

\bibitem[{\citenamefont{Silver et~al.}(2017)\citenamefont{Silver,
  Schrittwieser, Simonyan, Antonoglou, Huang, Guez, Hubert, Baker, Lai, Bolton
  et~al.}}]{AlphaGoZero}
\bibinfo{author}{\bibfnamefont{D.}~\bibnamefont{Silver}},
  \bibinfo{author}{\bibfnamefont{J.}~\bibnamefont{Schrittwieser}},
  \bibinfo{author}{\bibfnamefont{K.}~\bibnamefont{Simonyan}},
  \bibinfo{author}{\bibfnamefont{I.}~\bibnamefont{Antonoglou}},
  \bibinfo{author}{\bibfnamefont{A.}~\bibnamefont{Huang}},
  \bibinfo{author}{\bibfnamefont{A.}~\bibnamefont{Guez}},
  \bibinfo{author}{\bibfnamefont{T.}~\bibnamefont{Hubert}},
  \bibinfo{author}{\bibfnamefont{L.}~\bibnamefont{Baker}},
  \bibinfo{author}{\bibfnamefont{M.}~\bibnamefont{Lai}},
  \bibinfo{author}{\bibfnamefont{A.}~\bibnamefont{Bolton}},
  \bibnamefont{et~al.}, \bibinfo{journal}{Nature}
  \textbf{\bibinfo{volume}{550}}, \bibinfo{pages}{354} (\bibinfo{year}{2017}).

\bibitem[{\citenamefont{Herbrich et~al.}(2006)\citenamefont{Herbrich, Minka,
  and Graepel}}]{TrueSkill}
\bibinfo{author}{\bibfnamefont{R.}~\bibnamefont{Herbrich}},
  \bibinfo{author}{\bibfnamefont{T.}~\bibnamefont{Minka}}, \bibnamefont{and}
  \bibinfo{author}{\bibfnamefont{T.}~\bibnamefont{Graepel}}, pp.
  \bibinfo{pages}{569--576} (\bibinfo{year}{2006}),
  \urlprefix\url{http://dl.acm.org/citation.cfm?id=2976456.2976528}.

\end{thebibliography}

\appendix

\end{document}